# Pulsed neutron diffraction study of Zr-rich PZT

B. Noheda*, J.A. Gonzalo* and M. Hagen**.

*Dept. Física de Materiales, U. A.M. Cantoblanco, 28049-Madrid, Spain.
** Department of Physics, Keele University, Keele, Staffs, ST5 5BG, UK.

Pulsed neutron diffraction investigations have been performed in the ferroelectric PZT system ($PbZr_{1-x}Ti_xO_3$) doped with 1%wt. of $Nb_2O_5$, as a function of both temperature and composition. The study has been made in a wide range of temperatures encompassing the three known phases in Zr-rich PZT: ferroelectric low temperature (R3c), ferroelectric high temperature (R3m) and paraelectric (Pm3m). The combination of the temperature and the composition dependence of the structural parameters allowed the determination of the special relationship, recently pointed out by Corker et al., between the octahedral strain ($\zeta$) and the tilt angle ($\omega$) in the ferroelectric low temperature phase ($F_L$) of PZT. The strain-tilt coupling coefficient has been found to decrease linearly with Ti content and the composition at which $\zeta$ and $\omega$ de-couple has found to be of x≈ [1]0.30.

.

## 1 Introduction

PZT ($PbZr_{1-x}Ti_xO_3$) is a ferroelectric material widely used in industry. Its high piezoelectric response around the morphotropic phase boundary (x≈ 0.47) has given rise to a large amount of works in this range of compositions. Nevertheless its physical properties are still far from being fully understood. The Zr-rich part of the phase diagram has being comparatively less investigated. In this region the material undergoes two phase transitions: the high temperature ferroelectric ($F_H$) to paraelectric ($P_C$) transition, where the material goes from a rhombohedral phase (R3m) to a perovskite cubic phase; and a lower one between two ferroelectric phases ($F_L$ and $F_H$), being $F_L$ also rhombohedral with spatial group R3c. A high pyroelectric figure of merit is associated to this phase transition, which makes it also interesting from the point of view of applications.

The ferroelectric character of rhombohedral PZT is due to the cations displacement along the cubic [111] direction with respect to the oxygen planes. The low temperature phase transition ($F_L$) is driven by the tilts of the oxygen octahedra about [111] that double the unit cell and give rise to superlattice peaks, making necessary the use of neutrons to characterise it. This transition was firs investigated by neutron powder diffraction by Glazer et al.[1] for X=100·x= 10, and they already pointed out the unreasonable values of the anisotropic thermal factors resulting from the Rietveld refinement in rhombohedral PZT. Negative temperature factors are found often and the lead thermal ellipsoids are flattened perpendicular to [111]), indicating that the structural determination does not give a complete explanation of the data.

Although this point has not been yet completely clarified, in the last years Zr-rich PZT has attracted renewed interest, due to the evidence of short range ordering found by Viehland et al.[2] by electron microscopy, after getting some extra reflections incompatible with the R3c average symmetry. Work has been done by Teslic et al.[3] in order to obtain the local structure of rhombohedral PZT by analysing the Pair Distribution Function. A model considering the presence of localised regions with antiparallel cation displacements has been recently proposed by Ricote et al.[4] as an explanation for the extra reflections found by electron microscopy.

Very recently, using neutron diffraction data for several compositions at RT, Corker et al.[5] have shown that consistent Rietveld refinements with realistic anisotropic temperature factors can be obtained by considering random <100> Pb displacements superimposed to those along the [111] direction. The proposed Pb displacements are related by the same authors, to the existing coupling between the octahedral strain and the tilt angle in rhombohedral PZT, which shows anomalous behaviour for this system in comparison to other rhombohedral perovskites.

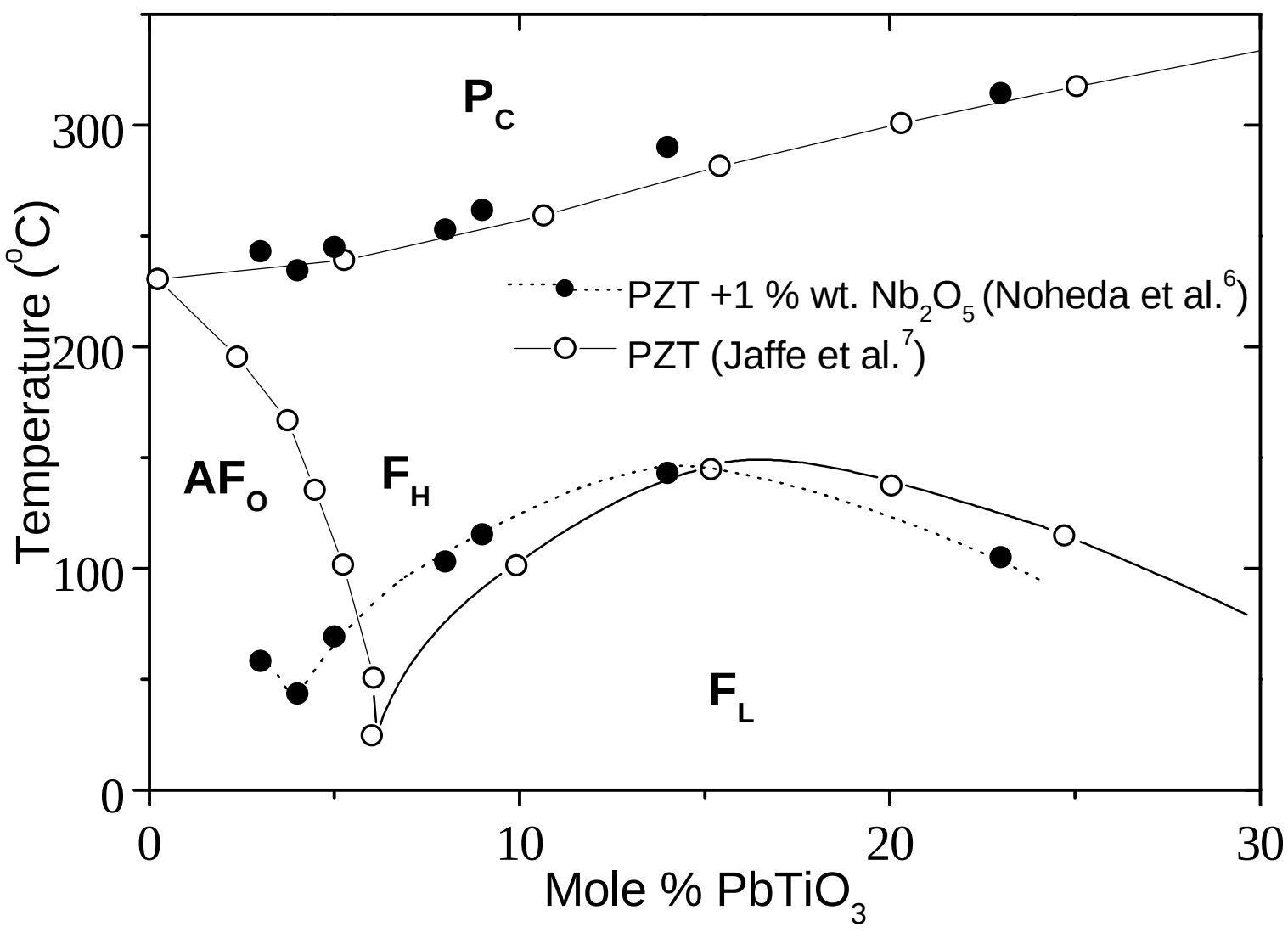

Figure 1. The phase diagram of PZT+1%wt. $Nb_2O_5$ in the rhombohedral region is shown together with the pure PZT phase diagram.

In this work the structural parameters have been obtained as a function of both composition and temperature in rhombohedral PZT by pulsed neutron powder diffraction. The data combining temperature and composition variation has allowed us to characterise the coupling between the octahedral strain (ζ) and the octahedral tilt (ω) and to obtain the previously unknown composition dependence of the coupling coefficient, which was found to be linear with the Ti content. The specific composition at which the de-coupling between ζ and ω occurs has been also obtained for our 1%wt. $Nb_2O_5$ doped PZT samples.

## 2 Experimental

The neutron measurements showed here were performed at the ISIS spallation neutron source at the EPSRC Rutherford Appleton Laboratory, UK, using the PRISMA spectrometer.

In this experiment we used PRISMA-II detector module, consisting on 16 detectors. Four detectors (with the analyser crystals removed) were used to perform diffraction measurements while the remaining twelve detectors (with analysers in place) were used to measure inelastic cuts through the powder spectrum. In this work we present only an elastic measurements analysis.

Nb-doped PZT ceramic cylindrical samples of three different composition, X=100·x= 4, 14 and 23, were measured at different temperatures through the $F_L$-$F_H$ and the $F_H$-$P_C$ phase transitions. Nb-doping was found to be very useful in reducing dielectric losses and increasing the pyroelectric coefficient. A small Nb amount, which occupies Zr/Ti sites, only modifies the transition temperatures, keeping unchanged the general characteristics of the PZT phase diagram. This can be seen in Figure 1, where the phase diagram of the 1% wt. Nb-doped PZT[6] is shown together with that of pure PZT[7].

The structure analysis has been done using the GSAS program (General Structure Analysis System, R.B. Dreele and A.C. Larson, 1996) for Rietveld refinement. Anisotropic temperature factors have been considered for all the atoms, and the Zr, Ti and Nb atoms have been treated together. Using the Suortti[8] empirical absorption function, which takes into account surface roughness effect, negative temperature factors are avoided in all the cases with the exception of X=23 and X= 14 at 200ºC and 250ºC, respectively. Convergence was achieved easily for X=4 but it became more difficult as the Ti content increased and, specially, at high temperatures (T> 200ºC).

Figure 2 shows a typical powder diffraction pattern, where intensity is plotted as a function of the neutrons time of flight (ToF), together with the Rietveld fit and the difference between both. The fit residuals, $R_p$, are of the order of 6%.

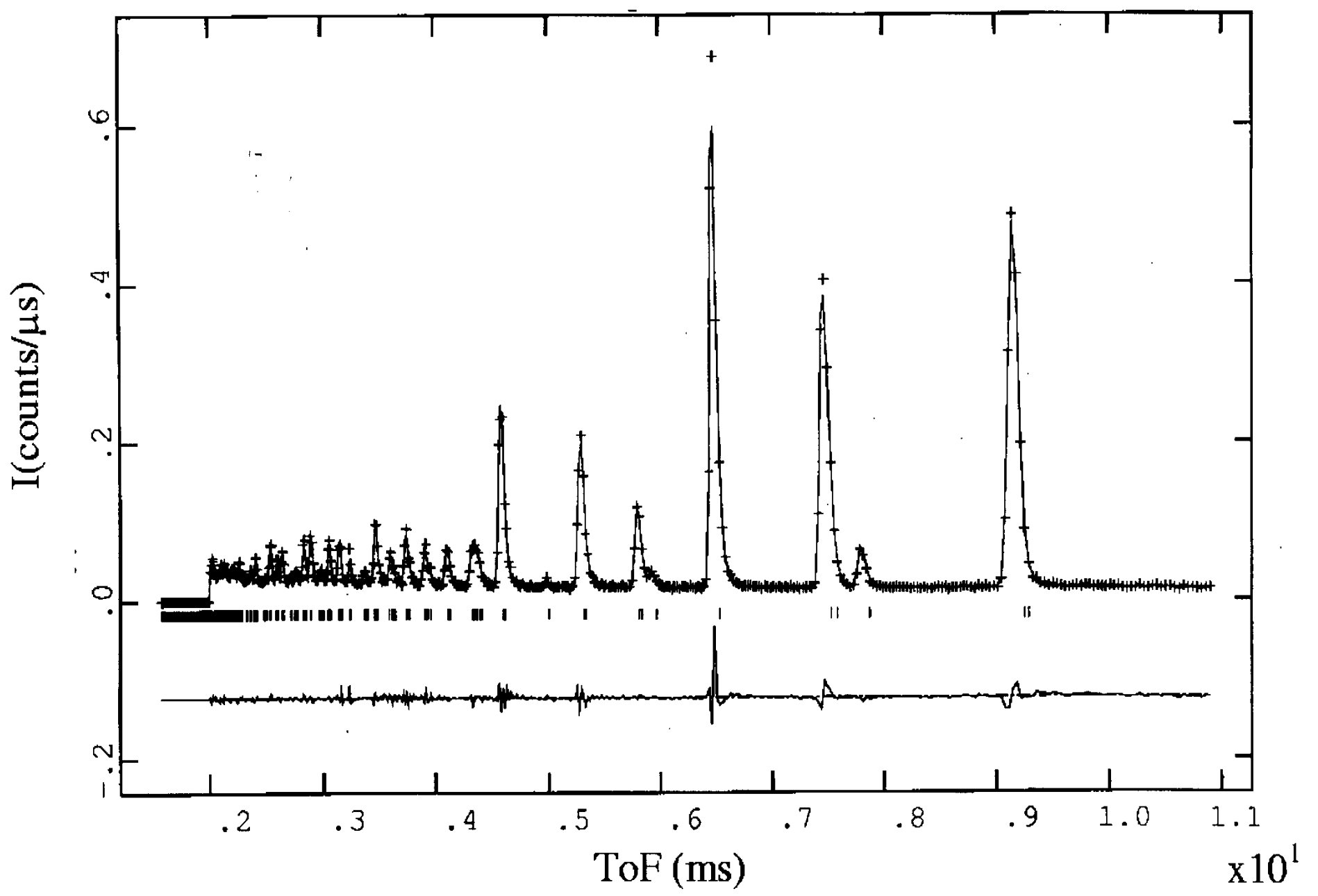

Figure 2. Observed (crosses) and calculated (line) neutron intensity profiles as a function of neutrons time of flight (ToF). The lower plot is the difference between the observed and the calculated profiles.

## 3 Results and discussion

Pb and Zr/Ti/Nb fractional shifts along the rhombohedral axis (s and t, respectively) are plotted in Figure 3 as a function of temperature for the three compositions studied. The dependence with temperature and composition, through the $F_L$-$F_H$ and $F_H$-$P_C$ transitions, agrees with that observed in the spontaneous polarisation measured for the same compositions[6]. As it can be seen, the $F_L$-$F_H$ transition is very well defined for X= 4 and it becomes more diffuse as the Ti content increases.

The structural parameter, *e*, being the order parameter driving the $F_L$-$F_H$ transition, is plotted in Figure 4 versus temperature through the $F_L$-$F_H$ transition temperature for the three mentioned compositions. *e* is related to the tilt angle, $\omega$, of the oxygen octahedra about the triad axis [111], through $\tan\omega = 4\sqrt{3}e$. It can be seen that the transition shows first order character for X=4 and the sharpness of the transition decreases with increasing X, as observed also in Figure 3. Thermal hysteresis measurements[6] in these samples indicate that there is an evolution from first to second order as Ti content increases, but this is masked by the increased diffuse character of the transition. This is probably due to Zr/Ti chemical disordering, that is, to the presence of Zr-rich and Ti-rich regions, which increases as we depart from the phase diagram at X= 0. Our refinement does not distinguish the individual positions of Zr and Ti atoms and the way these atoms are distributed in the lattice. Their difference becomes more evident as the Zr/Ti ratio approaches one. So it is not surprising that we obtain larger uncertainties in the Rietveld refinement with increasing X (see figure 4).

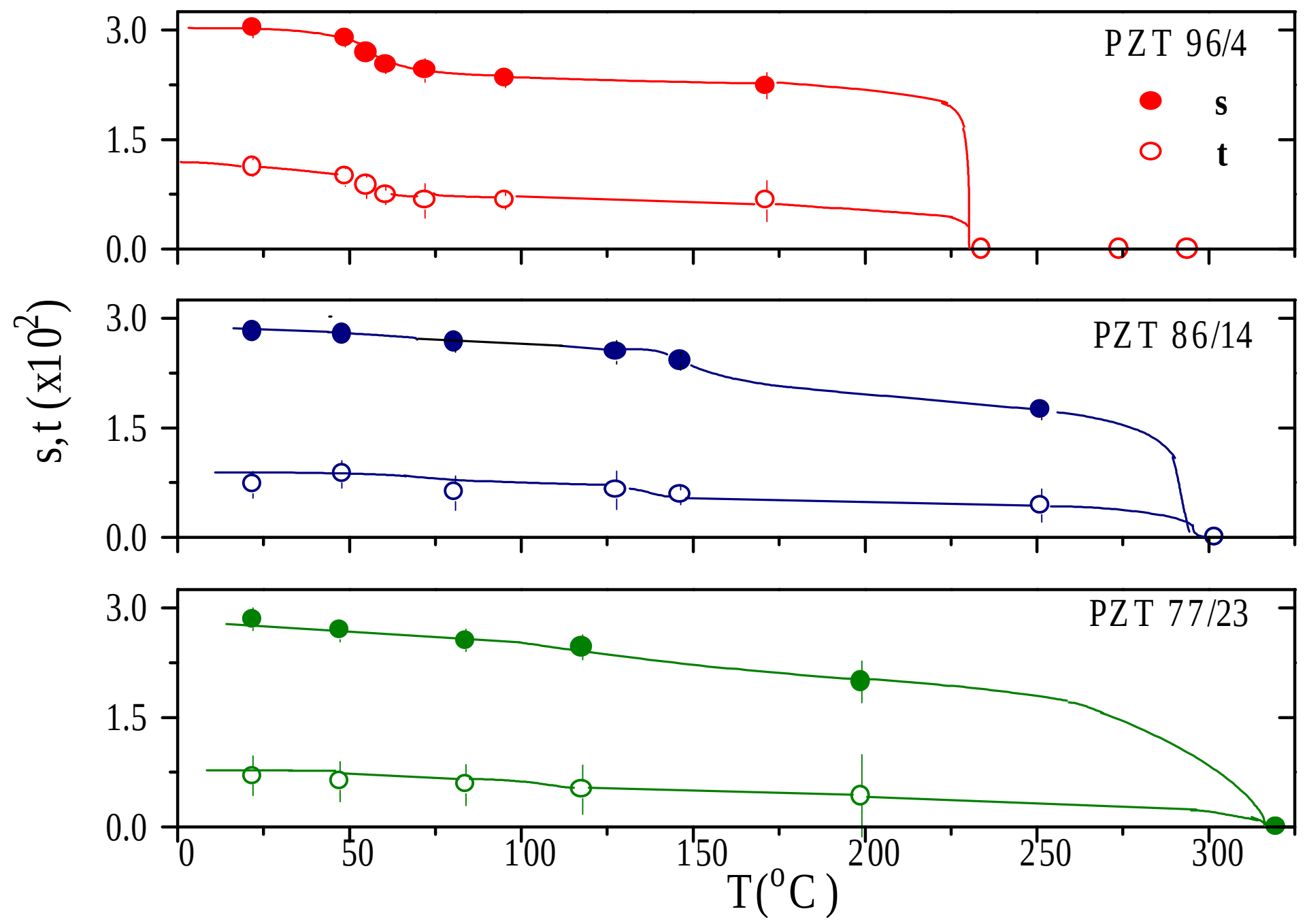

Figure 3. Pb and Zr/Ti/Nb shifts (s, t) along the polar axis versus temperature, for X= 4, 14 and 23. Lines are plotted taking into account the spontaneous polarisation measurements [6].

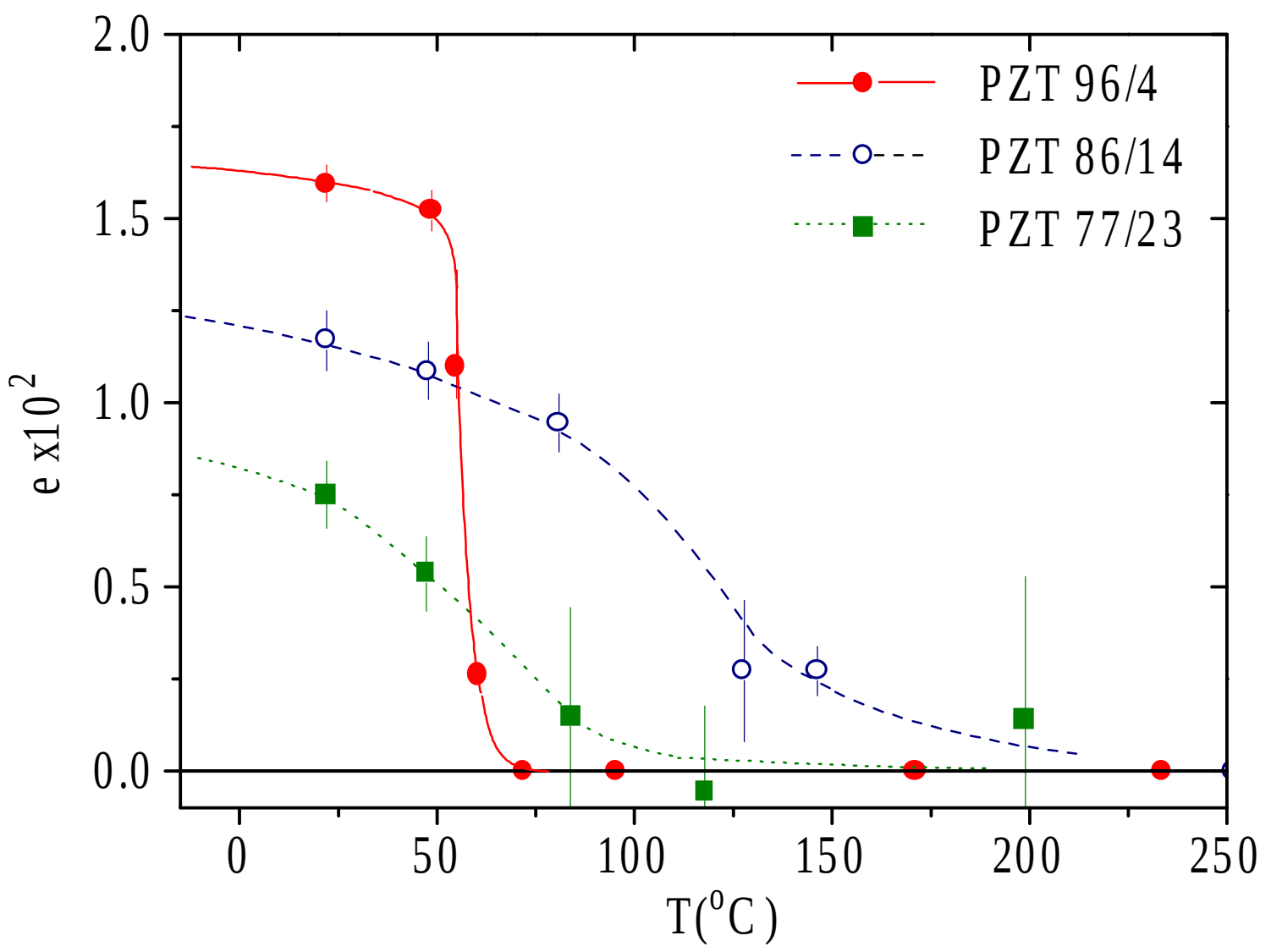

Figure 4. Structural parameter e, describing the octahedra tilt, versus temperature, for X= 4, 14 and 23. Note that the uncertainties for X= 23, at T≈ 84, 118 and 200 $^o$C, are very substantial, including e= 0 within the error bars.

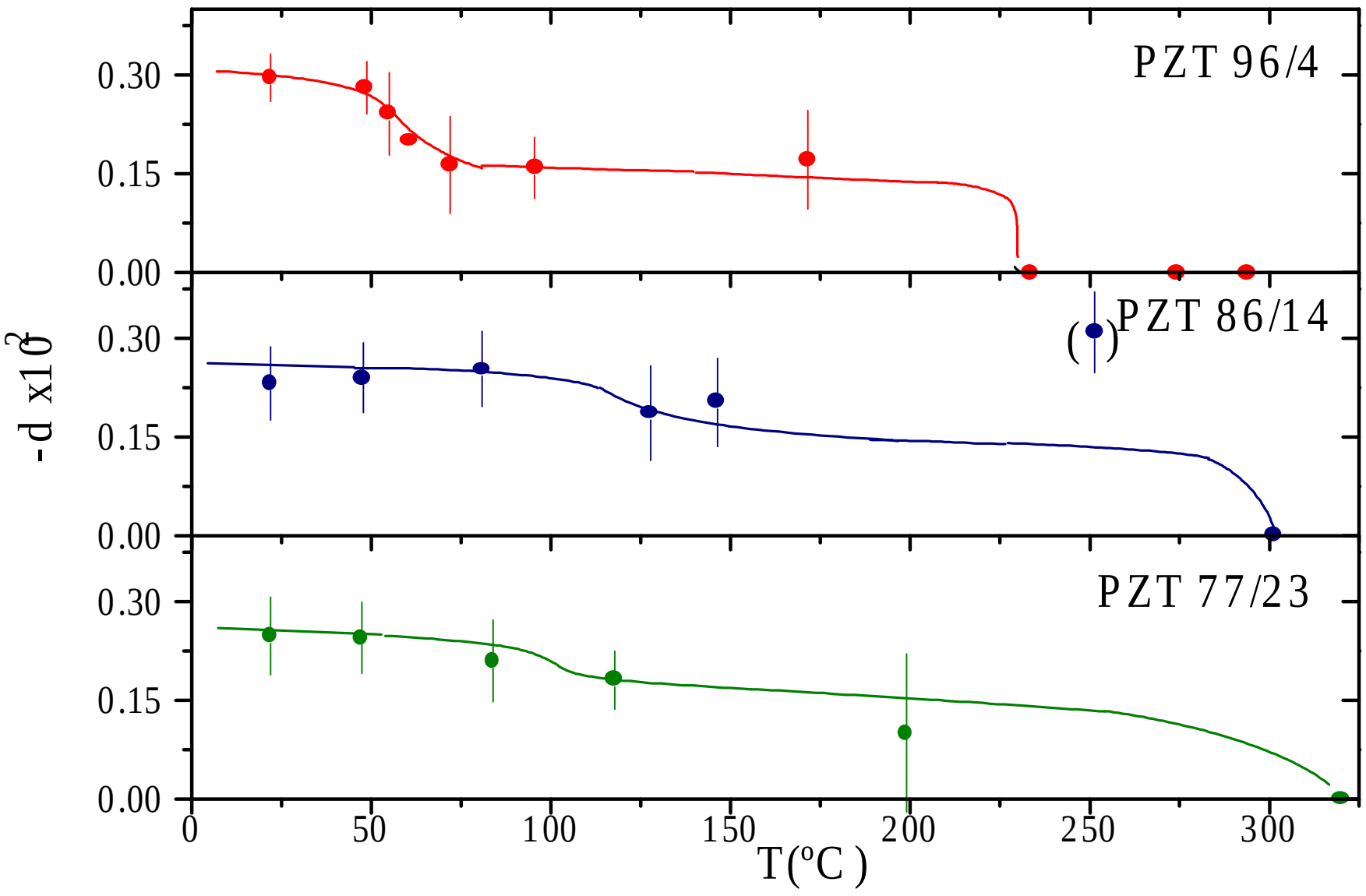

Fig 5. Minus the octahedral distortion, d, vs. temperature for X= 4, 14 and 23.

The octahedral distortion, d, describing the different size of the two oxygen triangles forming the octahedra, is plotted in Figure 5. The big jump in d occurring in the $F_L$-$F_H$ phase transition, allows a very good definition of the transition temperature, in spite of its small values and the sizeable standard deviations. As far as we know it is the first time that the octahedral distortion (d) is shown to have a definite behaviour across the $F_L$-$F_H$ phase transition.

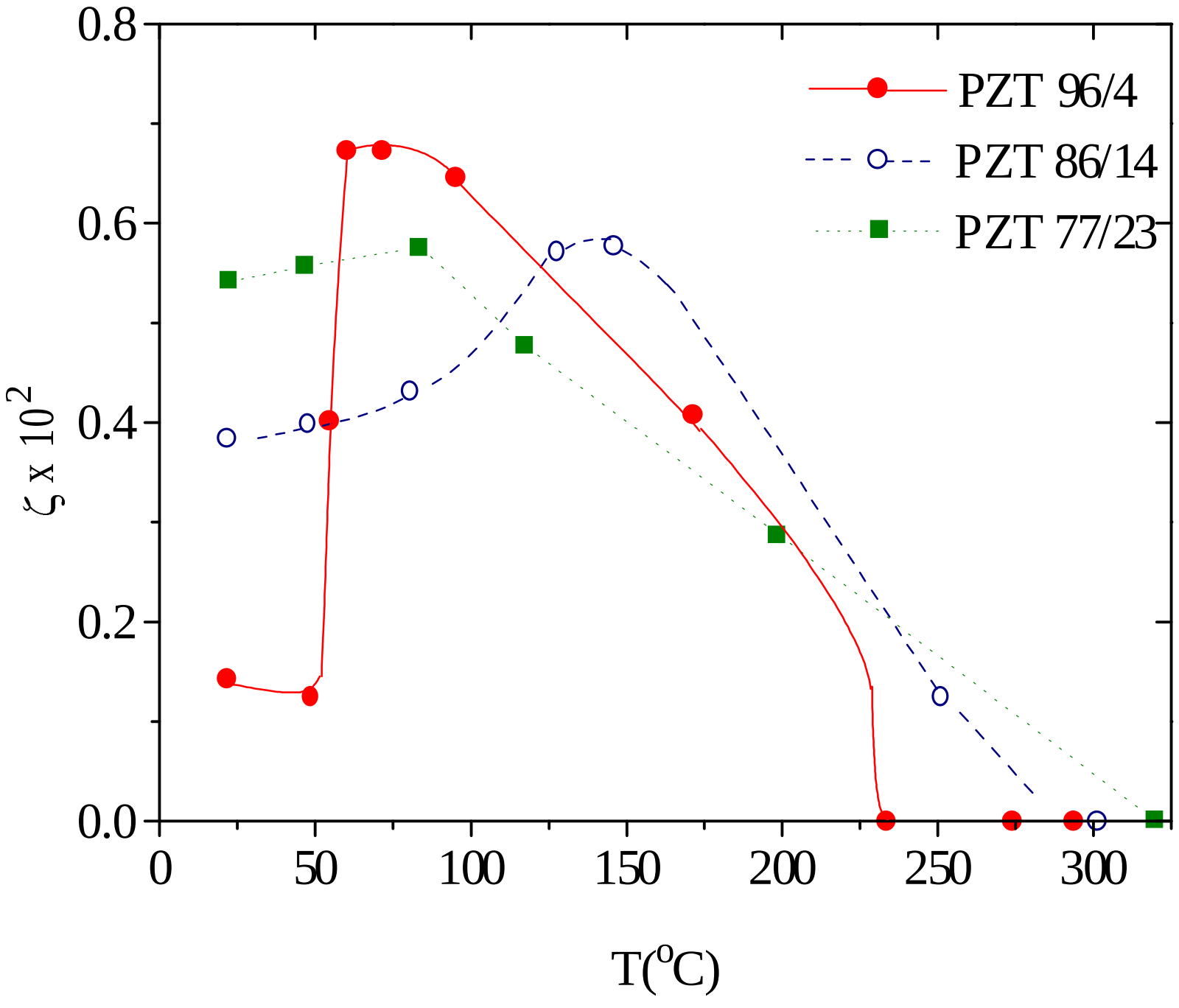

Figure 6. The octahedral strain vs. temperature for the three studied compositions.

The octahedral strain, defined as $\zeta = \cos\omega\ [c_h/(a_h\sqrt{6})] - 1$, where $a_h$ and $c_h$ are the hexagonal unit-cell parameters, is plotted in Figure 6 as a function of temperature for the compositions studied. The observed increment of the octahedral strain, $\zeta$, with temperature in the $F_L$ phase, which is simultaneous with the decrease of the octahedra tilt, $\omega$, has not been reported in other rhombohedral perovskites, and does not agree with previous geometrical considerations[9], as pointed out by Corker et al.[5]. The same authors propose an expansion of the Gibbs energy function in terms of both $\zeta$ and $\omega$ ($G = G_o + \frac{1}{2} a\cdot\omega^2 + \frac{1}{4} b\cdot\omega^4 + \frac{1}{6} c\cdot\omega^6 + d\cdot\zeta + \frac{1}{2} e\cdot\zeta^2 + f\ \omega^2\cdot\zeta$). Including the third order coupling between $\omega$ and $\zeta$, a linear relation:

$$\zeta = -d/e - (f/e)\ \omega^2 \equiv A(X) - B(X)\ \omega^2 \quad , \qquad (1)$$

is obtained, being A(X) and B(X) composition dependent coefficients.

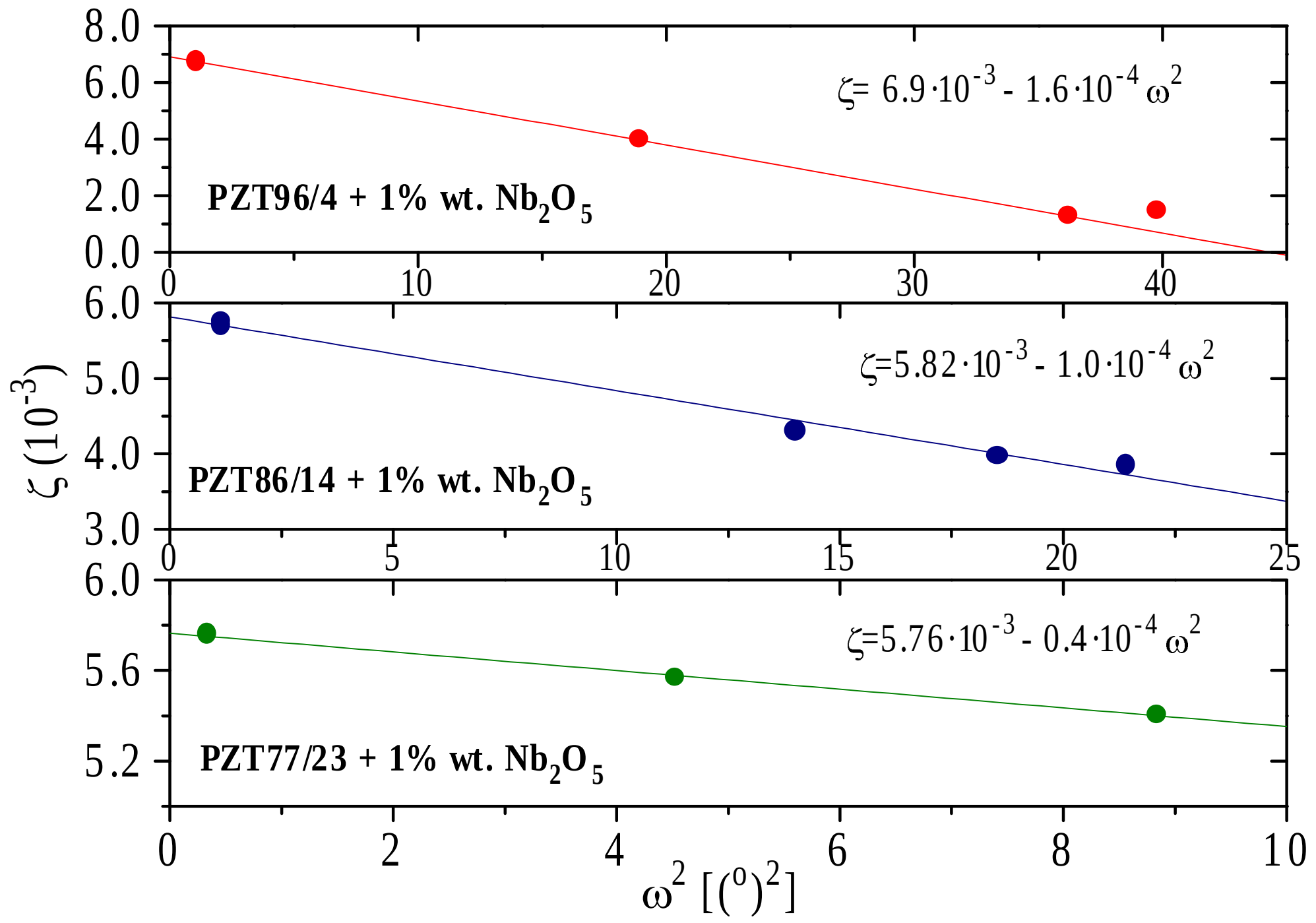

Figure 7. Octahedral strain vs. squared tilt angle for X= 4, 14 and 23. Lines are the linear fits.

Combining the temperature and the composition dependence of the structural parameters $\zeta$ and $\omega$ allows us to describe the existing octahedral strain-tilt coupling in rhombohedral PZT. Figure 7 shows the octahedral strain as a function of the squared tilt angle for X= 4, 14 and 23, indicating that equation (1) is fulfilled in the three cases. This behaviour distinguishes PZT from other rhombohedral perovskites, and seems to be related to with the existence of locally ordered regions with other symmetry than that of the average crystal structure[5].

Data in Figure 7 have been fitted for each composition to obtain the coupling coefficient B(X) of equation (1), which is proportional to the coupling factor, f in the Gibbs energy function. We see in Figure 8 that B(X) decreases linearly with increasing X. It also shows that there is a certain critical composition, $X \approx 30$ for PZT + 1%wt. $Nb_2O_5$, at which the octahedral strain-tilt decoupling takes place. This is in agreement with the disappearance of the superlattice reflections found by electron microscopy[2,4] close to the morphotropic phase boundary of PZT ($X \approx 47$).

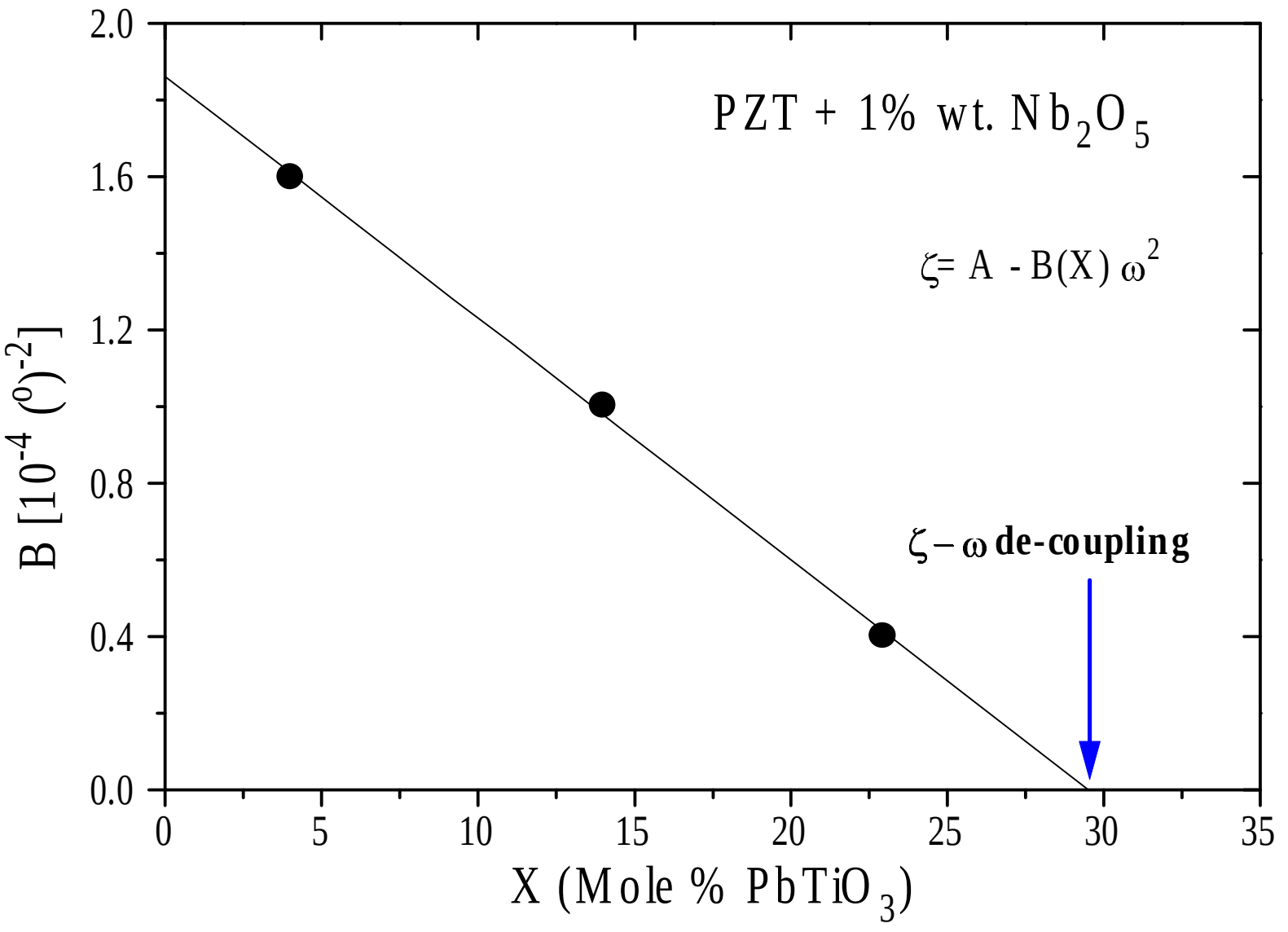

Figure 8. Coupling coefficient, B(X), as defined in equation (1), vs. X

In conclusion, a pronounced temperature dependence of the octahedral distortion, d, through the $F_L$-$F_H$ phase transition is reported for the first time, as far as we know. The coupling between the octahedral strain (ζ) and the octahedral tilt (ω) has been characterised in detail for rhombohedral PZT, using data at various compositions and at different temperatures through the $F_L$ phase. The coupling coefficient has been found to be linear dependent with composition. At a critical composition of X≈ 30 the coupling becomes negligible, a result which is in agreement with the compositional evolution of the extra-reflections found by electron microscopy, and confirms the direct relation between the coupling and the reported local structures.

**Acknowledgements**

Our thanks to Dr. Ning Duan and the Shanghai Institute of Ceramics for supplying the samples. Helpful comments by Prof. A.M. Glazer are gratefully acknowledged. We thank Martyn Bull and Steve Bourne for their help with the data collection and analysis. The financial support of CICyT (PB96-0037) and NATO (CGR-0037) is also acknowledged.

**References**

1. A. M. Glazer, S. Mabud, R. Clarke. *Acta Cryst.*1978, B34, 1060-1065.
2. D. Viehland, J.F. Li, Z. Xu. *J. Phys. Chem. Sol.* 1996, 57, 1545- 1554.
3. S.Teslic, T.Egami, D.Viehland. *J.Phys. Chem. Sol.*, 1996, 57, 1537-1543 .
4. J. Ricote, D.L. Corker, R.W. Whatmore, S.A. Impey, A.M. Glazer, J. Dec, K. Roleder. *J. Phys.:Condens. Matter* , 1998, 10, 1767- 1786.
5. D.L. Corker, A. M. Glazer, R.W. Whatmore, A. Stallard and F. Fauth. *J. Phys.:Condens. Matter* , 1998, 10, 6251-6269.
6. B.Noheda, N. Duan, N. Cereceda, J. A. Gonzalo. *J. Korean Phys. Soc.*, 1998, 32, 256-259.
7. B. Jaffe, W.R. Cook, H. Jaffe. *Piezoelectric Ceramics* (London:Academic, 1971)
8. Suortti. *J. Appl. Cryst.* 1972, 5, 325-330,.
9. H. D. Megaw, C.N.W. Darlington. *Acta Cryst. A***31**, 161 (1975).